\documentstyle[12pt]{article}
\begin{document}
\hfill \vbox{\hbox{UCLA/95/TEP/39  }
              \hbox{hep-th/9601082 }}
\begin{center}
{\Large \bf Exact Relation between Einstein and Quadratic Quantum 
Gravity}\footnote{Research supported in part by NSF Grant 
PHY89-15286} \\[2cm]
{\bf E. T. Tomboulis}\footnote{e-mail address: 
tomboulis@physics.ucla.edu} \\
{\em Department of Physics\\
University of California, Los Angeles \\
Los Angeles, California 90095-1547} \\[2cm]
\end{center}

\section*{Abstract}
We show the exact equality of the path integral of the general
renormalizable fourth order gravitational action to the path integral
of the Einstein action coupled to a massive spin-0 field and a massive
spin-2 ghost-like field with non-polynomial interactions.  The metric
in the Einstein version is a highly nonlinear function of the metric
in the quadratic version. Both massive excitations are unstable. The
respective cosmological constant terms in the two versions can be very
different. Some implications are briefly discussed. 

\vfill

\pagebreak
Gravitational actions including higher than linear powers of the
Riemann tensor are, for a variety of reasons, interesting both as
classical and quantum field theories.  They, also, arise generically
in ( four-dimensional) string one-loop effective actions (see eg ref
\cite{Ov}). The relation between the Einstein-Hilbert action and such
nonlinear extensions has been addressed by many authors. One of the
earliest results is apparently that of ref. \cite{H}.  It establishes
that the addition of a term quadratic in the curvature scalar is
classically equivalent to the minimal coupling of a scalar field to
the E-H action plus a scalar field potential. At the classical level,
the most complete considerations were given in \cite{Mag}, \cite{KJ},
where it was shown that the fourth order action involving $R^2$ and
$R^2_{\mu\nu}$ terms is equivalent to the E-H action with a different
metric and coupled to a symmetric rank-2 tensor 'matter'
field. Equivalence here means that the two actions lead to equivalent
equations of motion. These results were reproduced, in a somewhat
different formalism, in \cite{HOv}, where it was further demonstrated
that the content of the tensor field is presisely that of a massive
pure spin-2 ghost-like field, and a massive spin-0 field.

In this paper we consider the quantum theory. Our approach is
motivated by that of \cite{Mag}-\cite{KJ}, but, in the quantum
context, we will obtain a rather stronger result. We will derive the
exact, albeit formal, equality of the functional integral of the most
general quadratic action (eq. (\ref{l1}) below) to that of the
Einstein theory with a new metric, and coupled to a massive spin-2
ghost-like field and a massive spin-0 field with nonpolynomial
interactions.  The metric in the Einstein version turns out to be a
highly nonlinear function of the metric in the quadratic theory. It is
somewhat remarkable that such an exact transformation of the
functional integral, where, of course, one integrates over all metrics
without regard of the equations of motion, can be given in closed
form. This exact equivalence has various physical implications that we
briefly discuss below. The quadratic action is renormalizable, whereas
the Einstein action with matter is not (by power counting). Both
massive fields turn out to be unstable, the massive spin-2 ghost being
actually unstable at tree level. The cosmological constant in the
Einstein version can be very different from that in the quadratic
action. The same method can be used to examine the relation between
other gravitational theories, for example, theories involving 
arbitrary powers of the Ricci tensor. Here, however, we restrict 
ourselves to the quadratic action, the protoype for this type of 
transformation.

Our starting point is the path integral for the general fourth order
gravitational theory
\begin{equation}
Z = \int [Dg_{\mu\nu}] \exp\left( i\int d^4 x {\cal L}(g)\right)
\quad,\label{z1}
\end{equation}
where\footnote{We use metric signature (+ - - -), 
$R_{\mu\nu} = \partial_\nu
\Gamma^\lambda_{\mu\lambda} - \ldots $,
 and $\kappa^2 \equiv 16\pi G$, where  
$G$ is Newton's constant.}
\begin{equation}
  {\cal L}(g) = \sqrt{-g}\;[\: \frac{\gamma}{\kappa^2}R 
-aR_{\mu\nu}R^{\mu\nu}
+bR^2 +\frac{\Lambda}{\kappa^4}\: ] \quad .\label{l1}
\end{equation}
The inclusion of appropriate gauge fixing and associated ghost terms
does not affect the derivation below and need not be indicated
explicitly. We now introduce an auxiliary, non-propagating field
$\chi_{\mu\nu}$ (of mass dimension 2), and rewrite (\ref{z1})-
(\ref{l1}) in the form:
\begin{equation}
Z = \int [Dg_{\mu\nu}][D\chi_{\mu\nu}] C^{1/2} 
\exp\left( i\int d^4 x {\cal L}
(g,\chi)\right)\quad, \label{z2}
\end{equation}
where
\begin{eqnarray}
{\cal L}(g,\chi)= \sqrt{-g}\;[\: \frac{\gamma}{\kappa^2}R & - &
 a(R_{\mu\nu}R^{\mu\nu}-
\chi_{\mu\nu}g^{\mu\kappa}g^{\nu\lambda}\chi_{\kappa\lambda})\nonumber\\
 & + & b(R^2- (g^{\mu\nu}\chi_{\mu\nu})^2)
 -\frac{\lambda}{\kappa^2}g^{\mu\nu}\chi_{\mu\nu}\: ] \quad .\label{l2}
\end{eqnarray} 
In (\ref{l2})
\begin{equation}
\lambda^2= \Lambda (4b-a)\quad ,\label{nlam}
\end{equation}
and $C\equiv Det\left(\sqrt{-g}[a(g^{\mu\kappa}g^{\nu\lambda}
+g^{\mu\lambda}g^{\nu \kappa})/2
-bg^{\mu\nu}g^{\kappa\lambda}]\right)$ - such purely local
determinants, which are actually equal to unity in dimensional
regularization, can be absorbed in the definition of the uncoupled
measure $[Dg_{\mu\nu}]$. Integration over $\chi_{\mu\nu}$ gives, of
course, back (\ref{z1})-(\ref{l1}).

We now introduce the quantity 
\begin{equation}
{\cal R}_{\mu\nu}(g,\chi) \equiv R_{\mu\nu} + \chi_{\mu\nu}\;\;\;,
\qquad {\cal R}=g^{\mu\nu}{\cal R}_{\mu\nu}\quad , \label{sR}
\end{equation}
in terms of which one has 
\begin{eqnarray}
{\cal L}(g, \chi) & = & \sqrt{-g}\left[ \frac{\gamma}{\kappa^2}{\cal R} - 
a{\cal R}_{\mu\nu}{\cal R}^{\mu\nu} + b{\cal R}^2 \right]\nonumber\\
                &  & \quad - \: \sqrt{-g}\left[ \frac{1}{\kappa^2}
 ( \gamma + \lambda)g^{\mu\nu} - 2a{\cal R}^{\mu\nu} +
 2b{\cal R}g^{\mu\nu}\right] \chi_{\mu\nu} \nonumber \\
   & \equiv & \hat{{\cal L}}(g, {\cal R}) - \frac{1}{\kappa^2}
 {\cal F}^{\mu\nu}(g,\chi) \chi_{\mu\nu} \quad. \label{l3}   
\end{eqnarray}
In (\ref{l3}), we have set 
\begin{equation}
\frac{1}{\kappa^2}{\cal F}^{\mu\nu}(g,\chi) = 
 \sqrt{-g}\left[ \frac{1}{\kappa^2} ( \gamma + \lambda)g^{\mu\nu}
 - 2a{\cal R}^{\mu\nu} + 2b{\cal R}g^{\mu\nu}\right] \; . \label{sF}
\end{equation}
The r.h.s. of (\ref{sF}) defines a composite tensor density field
 ${\cal F}^{\mu\nu}$, which we want to use as a new metric field. To
 this end we insert unity in the integrand in (\ref{z2}) in the form
 of a $\delta$-function integration over a tensor density field
 $\hat{h}^{\mu\nu}$:
\begin{eqnarray}
Z = \int [Dg_{\mu\nu}][D\chi_{\mu\nu}] [D\hat{h}^{\mu\nu}] &  C^{1/2} &
\prod_{x,\mu,\nu}\delta \left[ \sqrt{-h}h^{\mu\nu} 
- {\cal F}^{\mu\nu}
(g,\chi)\right] \nonumber\\
  &  & \exp \left(i\int d^4x {\cal L}(g,\chi)\right)\; . \label{z3}
\end{eqnarray}
The tensor field $h^{\mu\nu}$ is, uniquely, defined through $
 \hat{h}^{\mu\nu} = \sqrt{-h}h^{\mu\nu} \;,\; h\equiv det\,h_{\mu\nu}
 $, with $h_{\mu\nu}$ the inverse of $h^{\mu\nu}$. (We could, of
 course, work throughout in terms of $\hat{h}^{\mu\nu}$, but it is
 more convenient to express the equations below in terms of
 $h^{\mu\nu}$.) Now, in the integrand in (\ref{z3}), the
 $\delta$-function allows one to write:
\begin{eqnarray}
{\cal L} & = & \hat{{\cal L}}(g,{\cal R}) -
 \frac{1}{\kappa^2}\sqrt{-h}h^{\mu\nu}\chi_{\mu\nu} \nonumber \\
   & = & \frac{1}{\kappa^2}\sqrt{-h}h^{\mu\nu}R_{\mu\nu}(h)
 +  \frac{1}{\kappa^2}\sqrt{-h}h^{\mu\nu}[R_{\mu\nu}(g)
 - R_{\mu\nu}(h)]\nonumber \\
  &    &  \qquad + \:\hat{{\cal L}}(g, {\cal R})
 - \frac{1}{\kappa^2} \sqrt{-h}h^{\mu\nu}{\cal R}_{\mu\nu}(g,\chi)
 \quad.\label{l4}
\end{eqnarray}
Furthermore, one can invert (\ref{sF}) to express ${\cal R}(g,\chi)$
in terms of $g_{\mu\nu}$ and ${\cal F}^{\mu\nu}(g,\chi) =
\sqrt{-h}h^{\mu\nu}$. The result is
\begin{eqnarray}
{\cal R}_{\mu\nu} = - \frac{1}{2}\frac{1}{\kappa^2}
 (4b -a)^{-1}(\gamma + \lambda)g_{\mu\nu} & - & \frac{1}{2a}
\frac{1}{\kappa^2}\frac{1}{\sqrt{-g}}\left[ \sqrt{-h}h^{\alpha
\beta}g_{\alpha\mu}g_{\beta\nu}\right. \nonumber \\
             &   & \left. - b(4b-a)^{-1}\sqrt{-h}h^{\alpha\beta}
g_{\alpha\beta}g_{\mu\nu} \right] \quad, \label{isR} 
\end{eqnarray}
and may be substituted in (\ref{l4}) to express ${\cal L}$ entirely in
terms of $g_{\mu\nu}$ and $h_{\mu\nu}$. One finds
\begin{eqnarray}
-{\cal V}(g,h) & \equiv & 
\hat{{\cal L}}(g, {\cal R}) - \frac{1}{\kappa^2}
 \sqrt{-h}h^{\mu\nu}{\cal R}_{\mu\nu}(g,\chi) \nonumber \\
               & = & -\frac{1}{\kappa^4}(\gamma^2 -
 \lambda^2) (4b-a)^{-1}\sqrt{-g} + \frac{1}{\kappa^4}\frac{\gamma}{2}
(4b-a)^{-1}\sqrt{-h}h^{\alpha\beta}
g_{\alpha\beta} \nonumber \\
              &  & \; +\frac{1}{\kappa^4}\frac{1}{4a}\sqrt{-h}\sqrt{\left(
\frac{-h}{-g}\right)}\left[ h^{\mu\nu}h^{\alpha\beta}g_{\mu\alpha}
g_{\nu\beta}- b(4b-a)^{-1}\left(h^{\alpha\beta}g_{\alpha\beta}
\right)^2\right]. \label{pot}
\end{eqnarray}
Also, using a standard formula of Riemannian geometry for the difference 
between the Ricci tensors of two different metrics $g_{\mu\nu}$ and
 $h_{\mu\nu}$, one obtains 
\begin{eqnarray}
{\cal L}_{kin} & \equiv & \frac{1}{\kappa^2}\sqrt{-h}h^{\mu\nu}
[ R_{\mu\nu}(g) - R_{\mu\nu}(h) ] \nonumber \\
              & = & \frac{1}{4}\frac{1}{\kappa^2}\sqrt{-h}h^{\mu\nu}
g^{\rho\sigma}g^{\kappa\lambda} \left[ 2\nabla_\kappa g_{\sigma\nu}
\nabla_\rho g_{\mu\lambda} - 2\nabla_\kappa g_{\sigma\nu}\nabla_\lambda
 g_{\mu\rho} \right.\nonumber \\
                 &  &  \quad + \left.\nabla_\kappa g_{\rho\sigma}
\nabla_\lambda g_{\mu\nu} - 2\nabla_\kappa g_{\rho\sigma}\nabla_\nu
 g_{\mu\lambda} + \nabla_\mu g_{\rho\lambda}\nabla_\nu
 g_{\sigma\kappa}\right] \nonumber\\
               &  &  + (\mbox{total divergence}) \quad , \label{kin}
\end{eqnarray}
with covariant derivatives $\nabla$ computed with the metric
$h_{\mu\nu}$.  From (\ref{l4}), (\ref{pot}) and (\ref{kin}) one sees
that the only remainng dependence on $\chi_{\mu\nu}$ in (\ref{z3}) is
in the argument of the $\delta$-function, where, from (\ref{sR}),
(\ref{sF}), $\chi_{\mu\nu}$ enters linearly. Assuming, as usual, that
one may interchange the order of integrations in the functional
integral (\ref{z3}), one may now integrate over $\chi_{\mu\nu}$ to
obtain
\begin{equation}
Z = \int [Dg_{\mu\nu}][D\hat{h}^{\mu\nu}] C^{-1/2}
 \exp\left( i\int d^4 x {\cal L}
(g,h)\right)\quad, \label{z4}
\end{equation}
where
\begin{equation}
{\cal L}(g,h) = \frac{1}{\kappa^2}\sqrt{-h}h^{\mu\nu}R_{\mu\nu}(h) + 
{\cal L}_{kin}(g,h) - {\cal V}(g,h) \quad , \label{l5}
\end{equation}
with ${\cal L}_{kin},\; {\cal V}$ given by (\ref{kin}),
(\ref{pot}). (\ref{l5}) is the Hilbert-Einstein Lagrangian for the
metric $h_{\mu\nu}$ coupled to the 'matter' field $g_{\mu\nu}$, and 
fully reproduces the results of \cite{Mag},\cite{KJ}, but now 
obtained as an exact transformation of the functional integral. To
extract the spin content of the rank-2 tensor field $g_{\mu\nu}$, we
decompose it into its trace and traceless components with respect to
$h_{\mu\nu}$:
\begin{equation}
\phi \equiv \frac{1}{4}\frac{1}{\kappa}h^{\mu\nu}g_{\mu\nu} \quad ,
 \qquad \frac{1}{\kappa}g_{\mu\nu} = \phi_{\mu\nu}
 + h_{\mu\nu}\phi \;\; . \label{dec}
\end{equation}
Then we write
\begin{equation}
g_{\mu\nu} = \kappa\phi I_\mu^{\ \beta} h_{\beta\nu} \:,
 \quad g^{\mu\nu}=
\kappa^{-1}\phi^{-1}h^{\mu\alpha}I^{-1}\ _\alpha^{\ \nu} \; ,
\quad g = \kappa^4
\phi^4Ih \;\;, \label{dec1}
\end{equation}
with
\begin{equation}
I_{\mu}^{\ \nu}\equiv \left[ \delta_\mu^\nu
 + \phi^{-1}\phi_{\mu\alpha}h^{\alpha\nu}\right] \;,
\quad I^{-1}\ _\mu^{\ \alpha} I_\alpha^{\ \nu} =\delta_\mu^\nu \;,
\quad I\equiv detI_\mu^{\ \nu}\; . \label{defI}
\end{equation}
Substituting the decomposition (\ref{dec}),(\ref{dec1}) in (\ref{pot})
one obtains (working now in terms of the 'matter' fields
$\phi_{\mu\nu}$,\ $\phi$ all indices are raised and lowered by the
metric $h_{\mu\nu}$):
\begin{eqnarray}
{\cal V} & = & -\frac{1}{\kappa^4}\frac{v_0}{\gamma}\Lambda\sqrt{-h} + 
   \frac{\gamma}{\kappa^2}v_0^{-1}(4b-a)^{-1}\sqrt{-h}\left(\phi -
 \frac{1}{\kappa}v_0\right)^2 \nonumber \\
        &   & +\frac{1}{\kappa^4}(4b-a)^{-1}\sqrt{-h}(I^{-1/2} -1)  +
 \frac{\gamma}{\kappa^2}v_0^{-1}(4b-a)^{-1}\sqrt{-h}(I^{1/2}
 - 1)\phi^2 \nonumber \\
       &   &- \frac{1}{\kappa^4}\frac{1}{4a}\sqrt{-h}I^{-1/2}\phi^{-2}
\phi_{\mu\nu}\phi^{\mu\nu} \; , \label{pot1}
\end{eqnarray}
with
\begin{equation}
v_0 \equiv \left[\gamma(1 - \lambda^2/\gamma^2)\right]^{-1}\; .
\end{equation} 
Note that, since 
\begin{equation}
I^{1/2} = 1 - \frac{1}{4}\phi^{-2}\phi_{\mu\nu}\phi^{\mu\nu}
 + \: \cdots\; , 
\label{expI}
\end{equation}
the last three terms in (\ref{pot1}) contain 
linear and higher powers of $\phi_{\mu\nu}\phi^{\mu\nu}$. Accordingly,
we shift
\begin{equation}
\phi = \frac{1}{\kappa}v_0 (1 + \kappa\varphi) \; ,\qquad
 \phi_{\mu\nu} = v_0 \varphi_{\mu\nu}\; ,
 \label{exp}
\end{equation}
so as to absorb the linear term in $\phi$ in ${\cal V}$ in
 (\ref{pot1}). ${\cal V}$ provides then the 
nonvanishng background value for $\phi$ which is necessary for 
consistency, since $g_{\mu\nu}$ must possess an inverse. We also
conveniently scaled the fluctuating fields $\varphi$,
$\varphi_{\mu\nu}$ so as to give standard normalization to their
 kinetic terms in ${\cal L}_{kin}$.  Thus  
\begin{equation}
g_{\mu\nu} = v_0h_{\mu\nu} + \kappa v_0(\varphi h_{\mu\nu} +
 \varphi_{\mu\nu})\; , \qquad \mbox{and} \qquad I_\mu^{\ \nu}=
 \left[\delta_\mu^{\ \nu} + 
\kappa\frac{\varphi_{\mu\alpha}h^{\alpha\nu}}{(1 +\kappa\varphi)}\right]
 \; .\label{exp1}
\end{equation}
Inserting (\ref{exp1}) in (\ref{z4}), (\ref{l5}), (\ref{pot1}),
 (\ref{kin}), 
we finally obtain: 
\begin{eqnarray}
Z & = & \int [D\varphi_{\mu\nu}][D\varphi][D\hat{h}^{\kappa\lambda}] 
C^{-1/2} \exp\left( i\int d^4 x {\cal L}(\{\varphi\},h)\right)\;
\; ,\label{z5}  \\
{\cal L}(\{\varphi\},h) &  = & \frac{1}{\kappa^2}\sqrt{-h}R(h)
 + \frac{1}{\kappa^4}\frac{v_0}{\gamma}\Lambda\sqrt{-h} \nonumber \\
  & & +  \frac{3}{2}\sqrt{-h}h^{\mu\nu}\nabla_\mu\varphi\nabla_\nu\varphi - 
\sqrt{-h}\nabla_\mu\varphi\nabla_\nu\varphi^{\mu\nu} - 
\frac{1}{2}m_0^2\sqrt{-h}\varphi^2 \nonumber \\
    &  & + \sqrt{-h}\left[ -\frac{1}{4}h^{\mu\nu}\nabla_\mu
\varphi_{\alpha\beta}\nabla_\nu\varphi^{\alpha\beta} +
 \frac{1}{2}\nabla_\nu\varphi^{\mu\alpha}\nabla_\mu\varphi_\alpha^{\ \nu}
\right] + \frac{1}{4}m^2\sqrt{-h}\varphi_{\mu\nu}\varphi^{\mu\nu} \nonumber\\
    &  & + {\cal L}_{kin}^I - {\cal V}^I \; , \label{l6} 
\end{eqnarray}
with 
\begin{equation}
m_0^2 =2\gamma v_0(4b-a)^{-1} \frac{1}{\kappa^2}\; , \qquad  m^2 = 
\left[ \frac{1}{a} + \frac{v_0}{\gamma}\Lambda \right]\frac{1}{\kappa^2}
\; .\label{mass}
\end{equation}
In (\ref{l6}) we wrote out explicitly only the parts of ${\cal
L}_{kin}$ and ${\cal V}$ that are bilinear in the fields $\varphi$,
$\varphi_{\mu\nu}$.  ${\cal L}_{kin}^I$, and ${\cal V}^I$ then denote
the non-polynomial interaction terms containing the trilinear and
higher couplings in $\varphi$, $\varphi_{\mu\nu}$ from the expansion
of ${\cal L}_{kin}$, and ${\cal V}$, resp., in powers of $\varphi$,
$\varphi_{\mu\nu}$ upon insertion of (\ref{exp1}) in (\ref{kin}),
(\ref{pot1}). (The unexpanded expressions (\ref{kin}), (\ref{pot1}),
with the replacements (\ref{exp}), (\ref{exp1}), give this
nonpolynomial Lagrangian in closed form.)

We have then obtained the exact transfomation of the path integral of
the general fourth order theory (\ref{z1})-(\ref{l1}) to the form
(\ref{z5})-(\ref{l6}).  Note that any convenient gauge-fixing term,
e.g. $ (\Box\partial_\mu g^{\mu\nu})^2$, plus associated FP-ghost
terms in (\ref{z1}) is tacitly carried along in the above derivation,
and is, at the very end, reexpressed through (\ref{exp1}) in terms of
$\varphi$, $\varphi_{\mu\nu}$ and $h_{\mu\nu}$. A convienient choice
of gauge in (\ref{z1})-(\ref{l1}) will, of course, not translate, in
general, into a convenient gauge for computations in terms of the
metric $h_{\mu\nu}$ in the theory in the form
(\ref{z5})-(\ref{l6}). According to the standard FP argument, however,
the path integral $Z$ is actually independent of the gauge-fixing term
choice, so, once the equivalence of (\ref{z1}) to (\ref{z5}) is
obtained in one gauge, one may change this gauge to any other in
(\ref{z5})-\ref{l6}), i.e. the equivalence holds independently of the
gauge choice. The same is, of course, true for any gauge invariant
correlation functions.

Now (\ref{l6}) is the Hilbert-Einstein action for the metric
$h_{\mu\nu}$ with a cosmological term and coupled to a massive spin-2
field $\varphi_{\mu\nu}$, and a massive spin-0 field $\varphi$. Note 
that the $\varphi_{\mu\nu}$ kinetic plus mass terms
in (\ref{l6}) are not in (the curved-space
generalization of) the canonical Pauli-Fierz form\footnote{To go over 
to that form, express (\ref{l6}) in terms of the traceful field 
\( \psi \equiv \varphi h_{\mu\nu} + \varphi_{\mu\nu}\). The bilinear 
parts of (\ref{l6}) are then precisely the massive spin-2 action of 
ref. \cite{S}, which, as shown there, can be brought to the 
Pauli-Fierz mass form by a somewhat different field decomposition 
of $\psi_{\mu\nu}$. The resulting scalar field is a mixture 
of our scalar field and $\nabla_\mu\nabla_\nu\varphi^{\mu\nu}$. 
Note that this affects the definition of 
the scalar mass term. We do not discuss such alternative formulations 
since they are not pertinent to our main point here.}. But they are an 
equally good formulation of the standard Fierz description of a pure 
spin-2 field, i.e. traceless
$\varphi_{\mu\nu}$, and $\nabla_\mu \varphi^\mu_{\ \nu} $ obeying a 
constraint equation, which follows from the equations of motion, and 
involves only {\it first} derivatives of the fields 
\(\varphi_{\mu\nu} , \varphi, h_{\mu\nu} \)\footnote{The tracelessness of 
$\varphi_{\mu\nu}$
must be remembered when performing the variation.} \cite{F}. The 
non-polynomial interaction terms ensure full gauge
invariance and continued absence of a spin-1 component beyond the
(curved-space) linearized approximation. The lagrangian (\ref{l6})
thus provides a realization of a complete, consistent coupling of a
massive spin-2 field to a gravitational background, a noteworthy fact
(\cite{O}, \cite{HOv}).

The spin-2 and spin-0 field kinetic terms in (\ref{l6}) come with
opposite signs. The overall sign is set by the (Newtonian limit of
the) EH term, which makes the spin-2 have the wrong sign, ie be a
ghost-like field. We thus find, in the full non-linear theory, the
same field content as in the linearized approximation to (\ref{l1})
\cite{S}. Note, however, that in the full theory this can only be
achieved by introducing a new metric field which is a highly nonlinear
function of the metric in (\ref{l1}); the exact relation between
(\ref{l1}) and (\ref{l6}) is non-perturbative.

It is natural to view (\ref{l1}) as the theory formulated in terms of
field variables suitable for the UV regime (energies higher than
Planck scale energy), and (\ref{l6}) as the theory in terms of
variables appropriate to the IR region (energies at or below Planck
scale). Indeed, recall that (\ref{l1}) is a renormalizable \cite{S},
and in fact asympotically free (in the coupling $ \alpha^2 \equiv
1/a$) lagrangian \cite{TT}, \cite{FT}, \cite{A}. Its loop perturbative
expansion is, therefore, applicable in the deep UV region. (\ref{l6}),
on the other hand, has, by power counting, the usual
non-renormalizable behavior of the Einstein theory coupled to matter.

It is not clear, in the absence of explicit computations, how the
renormalizability of (\ref{l1}) appears in (\ref{l6}). The 'matter'
fields circulating in loops must serve as regulators. Still, since the
metric in (\ref{l6}) is a highly nonlinear composite field in terms of
that in (\ref{l1}), simple direct order-by-order cancellation of the
nonrenormalizable divergences in the loop expansion of (\ref{l6})
presumably does not occur. Rather, one expects that the divergences
are cancelled on mass-shell upon nonlinear shifts of field variables, 
and/or resummation of appropriate 
subclasses of graphs. In any case, we seem to have an interesting
example of the equivalence of a renormalizable to a
(by power counting) non-renormalizable lagrangian.

The Einstein version (\ref{l6}) separates out the massless graviton,
which dominates at large distance scales, from the massive fields, and
hence is suitable for consideration of the IR regime. It is here that
the difficult dynamical issues of the S-matrix asymptotic states and
unitarity become relevant. At tree level, the massive fields have
masses naturally of the order of the Planck mass. Now, the
asymptotically free coupling $\alpha$ grows large in the IR, and
indeed tends to diverge below the Planck scale. Since the mass of the
spin-2 particle grows with it, this leads to the possibility that this
particle disappears from the spectrum at large distances; ie there is
``confinement'' of the massive spin-2 ghost-like excitation, as has
often been suggested in the literature. In any event, at the very
least, the following situation should apply. With $\alpha$ large, the
Compton wavelength of the massive spin-2 becomes comparable or less
than its Schwarzschild radius. This suggests that, by well-known
results \cite{Y}, already at the {\it classical} level, collapse of
the surrounding spacetime and formation of a trapped surface must
occur. In such a case, the usual expansion about the tree level
description of a bare particle propagating on some given background is
clearly not meaningful. Rather, the 0-th order approximation must
already include enough interaction effects to correctly describe the
appearance of such a highly dressed object, essentially a mini black
hole.  An even more basic question is that of the stability of
these massive particles, ie whether they can appear in the true
asymtotic states at all.  Inspection of the interaction terms in the
Lagrangian (\ref{l6}) shows that {\it both the spin-2 and the spin-0
particles are unstable}: there are trilinear vertices in ${\cal
L}_{kin}$ that, for $b \geq 9a/4$, allow the tree-level decay of the
$\varphi_{\mu\nu}$-particle into two $\varphi$-particles (plus
gravitons); and both particles can decay into gravitons through
radiative loop corrections. Now the S-matrix can, strictly speaking,
be defined only between in- and out-states containing solely stable
particles. In the standard field theoretic treatment of unstable
particles \cite{V}, this S-matrix connecting stable particles only is
constructed in terms of complete, dressed propagators for the unstable
particles; and can then be shown \cite{V} to be unitary and
causal. This formalism must be applied here too. In fact, the spin-2
particle appears as the simplest example of an unstable particle, ie
decaying at tree level (so it {\it has} to be treated in terms of
dressed propagators), except for the fact that its bare propagator has
negative residue. In this connection it might be also useful to recall
that, in the quantum theory, a negative residue can be traded for
negative energy flowing through the propagator. But in a gravitational
field there is no invariant meaning of local energy density.  So there
is nothing immediately inconsistent in allowing localized negative
energies, corresponding to the occurance of an unstable excitation, if
they do not affect the asymptotic values of the fields.

Another noteworthy feature of the equivalence of (\ref{l1}) to (\ref{l6}) 
is the relation between their respective cosmological terms. Both vanish 
if $\Lambda$ is fine-tuned to zero. 
For $\Lambda \neq 0$, however, the cosmological constant $v_0\Lambda/\gamma$ 
in (\ref{l6}) can be very different from $\Lambda$. If, in particular, the 
couplings run appropriately in the IR, it can tend to zero at large scales, 
even for large values of $\Lambda$. It is easily seen that there is more 
than one senario for the renormalization group flows of the couplings 
\begin{math} \gamma,\: a,\: b,\: \Lambda \end{math}
that could lead to this behavior. The asymptotic freedom of $1/a$ is
firmly established, but the present state of the computations of the
renormalization of the other couplings, \cite{FT}, \cite{A}, does not
yet allow one to draw any definitive conclusions.  More investigation
is needed to ascertain if this interesting possibility is actually
realized.

The method presented here can be used to examine the relation between other 
gravitational theories. One obvious application is to the supersymmetric 
version of (\ref{l1}) (cp ref \cite{Ov}). A more intriguing case 
is that of more general theories involving arbitrary polynomials 
in the Ricci tensor. Results will be presented elsewhere.

\end{document}